\documentclass{aa}
\usepackage{graphicx}
\usepackage{natbib}
\bibpunct{(}{)}{;}{a}{}{,}
\usepackage{xspace}
\usepackage[normalem]{ulem}

\newcommand{\micron}{\,$\mu$m }

\newcommand{\Msun}{M\ensuremath{_{\odot}}\xspace}
\newcommand{\ten}[1]{10\ensuremath{^{#1}}\xspace}

\begin{document}
   \title{Carbonaceous molecules in the oxygen-rich circumstellar environment of binary post-AGB stars:
   \subtitle{C$_{60}$ fullerenes and polycyclic aromatic hydrocarbons.}
\thanks{This research has made use of the SIMBAD database,
operated at CDS, Strasbourg, France.
Based on observations made with the SPITZER Space Telescope (program id 3274, 50092), which is operated 
by the Jet Propulsion Laboratory, California Institute of Technology under a contract with NASA.}}
	
   \author{
          C. Gielen\inst{1,2}\fnmsep 
          \thanks{Postdoctoral Fellow of the Fund for Scientific Research, Flanders}
     \and
     J. Cami \inst{3,4}
    \and
    J. Bouwman \inst{2}
    \and
    E. Peeters \inst{3,4}
    \and
    M. Min \inst{5}
          }

%  \offprints{C. Gielen}

   \institute{Instituut voor Sterrenkunde,
              Katholieke Universiteit Leuven, Celestijnenlaan 200D, 3001 Leuven, Belgium \\
	      \email{clio.gielen@ster.kuleuven.be}
	\and
	  Max Planck Institut f\"{u}r Astronomie, K\"{o}nigstuhl 17, 69117 Heidelberg, Germany %2
     \and
	Department of Physics and Astronomy, University of Western Ontario, London, Ontario N6A 3K7, Canada 
	\and
	SETI Institute, 189 Bernardo Ave, Suite 100, Mountain View, CA 94043, USA 
	\and
	Astronomical Institute, University of Utrecht, PO Box 80000, 3508 TA Utrecht, The Netherlands
	}

   \date{Received ; accepted }

% \abstract{}{}{}{}{}
% 5 {} token are mandatory
  \abstract
{The circumstellar environment of evolved stars is generally
    rich in molecular gas and dust. Typically, the entire environment
    is either oxygen-rich or carbon-rich, depending on the evolution of the central star. }
{ In this paper we discuss three evolved disc sources with evidence of atypical emission lines
in their infrared spectra. The stars were taken from a larger sample of post-AGB binaries for which we
have Spitzer infrared spectra, characterised by the presence of a stable oxygen-rich circumbinary disc. Our previous studies have shown that the infrared spectra of post-AGB disc sources are dominated by silicate dust emission, often with an extremely high crystallinity fraction. However, the three sources described here are selected because they show a peculiar molecular chemistry.}
   {Using Spitzer infrared spectroscopy, we study in detail the
    peculiar mineralogy of the three sample stars. Using the observed
     emission features, we identify the different observed dust,
     molecular and gas species.  }
   {The infrared spectra show emission features due to various
       oxygen-rich dust components, as well as CO$_2$ gas. All
       three sources show the strong infrared bands generally ascribed
       to polycyclic aromatic hydrocarbons. Furthermore, two sample
       sources show C$_{60}$ fullerene bands. }
{Even though the majority of  post-AGB disc sources are dominated by silicate dust in their circumstellar environment, we do find evidence that, for some sources
   at least, additional processing must occur to explain the presence of large carbonaceous molecules. 
   There is evidence that some of these sources are still oxygen-rich, which
       makes the detection of these molecules even more
       surprising.}
   \titlerunning{Carbonaceous molecules in the O-rich CE of binary post-AGB stars.}
   \maketitle
%
%________________________________________________________________

\section{Introduction}

The chemistry of the circumstellar environment (CE) of evolved stars
can typically be described as either oxygen-rich or carbon-rich, and
reflects the evolution of the central object along the
Hertzsprung-Russell diagram.  Stars on the Asymptotic Giant Branch
(AGB) start as oxygen-rich stars, but their carbon abundance rises
during the evolution on the AGB, since the carbon produced in the core
is dredged-up to the stellar photosphere.  If this dredge-up process
repeats enough times, carbon becomes more abundant than oxygen:
the star has become a carbon star. The main parameters determining the
mass of carbon dredged up are the initial mass of the
central star and its metallicity \citep[see][for a detailed
  description of AGB evolution]{herwig05}.

Depending on the C/O ratio of the central star, its CE will be either
oxygen-rich or carbon-rich.  The less abundant species will be locked
in carbon monoxide (CO), which is an extremely stable molecule and one
of the first to be formed. If the central star is oxygen-rich, the CE
is typically characterised by gas and dust features such as from OH, CO$_2$, H$_2$O, oxides,
and silicates \citep[e.g.][]{justtanont96,waters96,molster02a,gielen11}.  
For some oxygen-rich sources, the presence of carbon-rich molecules
can be explained by non-equilibrium effects, such as shocks \citep{nejad88,duari99,cherchneff06}.
For example, S-type stars, which are undergoing the transition from
oxygen-rich to carbon-rich and thus have C/O$\sim 1$, do not yet
form carbon-rich dust. However, due to non-equilibrium chemistry effects
carbon-rich molecules, such as PAHs, HCN, and C$_2$H$_2$ can be seen \citep{hony09,smolders10}.
In the carbon-rich case,
the excess carbon is used to form, for example, CN, C$_2$,
C$_2$H$_2$, CH$_4$, polycyclic aromatic hydrocarbons (PAHs), and SiC.
\citep[e.g.][]{bakker97,speck97,peeters02,hony03,matsuura06,volk11}. 
Recently, even more complex carbon molecules,
namely C$_{60}$ and C$_{70}$ fullerenes, were detected in the CE of
evolved stars
\citep{cami10,garciahernandez10,garciahernandez11,zhang11}.

In some cases, a strong mixed chemistry is seen in the CE of evolved objects
\citep{szczerba07b}.  
One of these is a particular class of AGB stars,
the silicate carbon stars, where the infrared spectra show
features of both C-rich species and O-rich silicates
\citep{yamamura00,molster01,garciahernandez06}.  Also some post-AGB
stars and proto-planetary nebulae (pPNe) show features of both PAHs
and silicates in their spectra
\citep{beintema96,waters98,matsuura04,cerrigone09,guzmanramirez11}.
For all these objects the dual chemistry is usually a result of the
evolution of the central star.  The O-rich matter was expelled when
the star was still oxygen-rich, followed by a transition from C/O$ <
1$ to C/O$ >1$ in the stellar photosphere. Subsequent outflow of
material will then be carbon-rich.  This effect is even more
pronounced if the O-rich material is stored in a stable disc around
the central star, as is the case for the silicate carbon stars
and several of the mixed-chemistry pPNe
\citep{waters98,deroo07a,cerrigone09}. 

In this paper we look in more detail to a sample of three post-AGB disc sources.
These sources are part of the larger sample of 57 post-AGB disc sources with Spitzer infrared spectra,
as discussed in \citet{gielen11}. In \citet{gielen11} we find that the circumstellar environment
of these post-AGB disc sources is dominated by emission of silicate dust.
However, four sources do show features attributed to carbonaceous molecules, in the form of PAHs and C$_{60}$ fullerenes, 
besides the silicate emission.
The carbonaceous content of one of these sources has already been discussed in \citet{gielen09}. In this paper we focus
on the remaining three stars. 

Emission of PAHs is
already observed in several post-AGB sources
\citep[e.g.][]{beintema96,cami01,peeters02,molster02b,matsuura04,gielen09}, where most of the
central stars appear carbon-rich.
There is evidence that the central stars from our sample are oxygen-rich, which would make the
detection of carbonaceous molecules even more surprising.
Moreover, this is the first time fullerenes are detected in O-rich post-AGB sources.  

\begin{table*}
\caption{The name, equatorial coordinates $\alpha$ and $\delta$(J2000), spectral type, effective
temperature $T_{\rm eff}$, surface gravity $\log g$ and metallicity [Fe/H] of our sample stars.
For the model parameters we refer to \citet{waelkens91b,deruyter06,hrivnak08}. 
The spectral type for IRAS\,13258 was determined from optical spectroscopy by Thomas Lloyd Evans (private communication).
Also given is the orbital period \citep[see references in][]{deruyter06,gielen07} and the C/O ratio \citep{luck84,bond87,waelkens91b,vanwinckel92,hrivnak08}}
\label{sterren}
\centering
\begin{tabular}{lrrcccrrrcc}
\hline \hline
Name & $\alpha$ (J2000) & $\delta$ (J2000)  & $T_{\rm eff}$ & $\log g$ & [Fe/H] & $P_{\rm orbit}$ & C/O\\ 
  & (h m s) & ($^\circ$ ' '')  & (K) & (cgs) & & (days) &\\
\hline
\object{IRAS\,06338+5333} & 06 37 52.4 & $+$53 31 02 & 6250 & 1.0 & -1.6 & 600 & 0.7\\
\object{IRAS\,13258-8103} & 13 31 07.1 & $-$81 18 30 & F4Ib-G0Ib & & & & & & &\\ 
\object{HD\,52961}    & 07 03 39.6 & $+$10 46 13  & 6000 & 0.5 & -4.8   & 1310 & 0.7 \\ 
\hline                                                                             
\end{tabular}
\end{table*}

The outline of the paper is as follows: in Section~\ref{sect_progstars} we present the programme stars and
observation details. All sample stars show features due to O-rich dust species, 
which are discussed in Sect.~\ref{sect_silicates}.
O-rich CO$_2$ gas emission is also seen in a few sources, and discussed in Sect.~\ref{sect_co2}. 
Clear emission of PAHs can be seen in the $6-12$\micron region of the observed infrared spectra,
as can be seen in Sect.~\ref{sect_pahs}. In Sect.~\ref{sect_c60} we report on the detection of the C$_{60}$ fullerenes and
derive some basic parameters such as the excitation temperature and mass.
We end with discussions and conclusions in Sects.~\ref{sect_discussion} and \ref{sect_conclusions}.

\section{Programme stars and observations}
\label{sect_progstars}

The binary post-AGB stars discussed here are those sources from the larger
Spitzer sample of post-AGB disc candidates \citep{gielen11}, showing features of carbonaceous molecules
besides the dominant silicate emission seen in these stars. The sample of disc candidates was
selected on the basis of their infrared colours, as discussed in \citet{deruyter06} and \citet{gielen11}
These binary post-AGB sources are characterised by the presence
of a stable O-rich dust disc, with strong features of amorphous and
crystalline silicates \citep{gielen08,gielen11}.  
These post-AGB disc sources typically have an O-rich stellar photosphere,
with no evidence for strong third dredge-up on the AGB,
as determined, for example, by the very low C/O ratio ($<< 1$), low $^{12}$C/$^{13}$C ratio, and the lack of s-process enhancement \citep[e.g.][]{gonzalez97a,vanwinckel03,maas05,reyniers07a,gielen09}.
This is surprising since several stars have initial masses or luminosities
high enough to evolve to a carbon star \citep{gielen09,gielen11}.
Apparently, the AGB evolution of these stars was shortcut,
probably under the influence of strong binary interaction.

\begin{figure}
\vspace{0cm}
\hspace{-0.7cm}
\resizebox{10cm}{!}{ \includegraphics{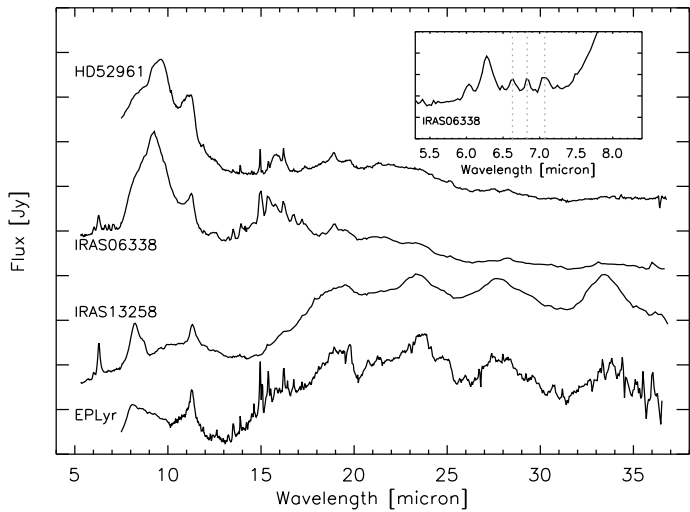}}
\caption{The Spitzer-IRS high- and low-resolution spectra of our three sample stars IRAS\,13258, IRAS\,06338 and HD\,52961,
together with the spectrum of EP\,Lyr, which shows similar emission features.
The spectra are scaled and offset for comparison. The inset plot shows the 7\,$\mu$m region of IRAS\,06338 and 
IRAS\,13258. The dotted gray lines mark the observed features at 6.63, 6.83 and 7.07\,$\mu$m, 
probably due to small PAH, aliphatic hydrocarbons, and/or hydrogenated amorphous carbon emission.}
\label{3stars}%
\end{figure}

All stars were observed with the Spitzer-IRS spectrograph, in high- or low-resolution modes \citep[resolution of $\sim 600$ and $\sim 160$ respectively][]{werner04,houck04}.
Exposure times were chosen to achieve a S/N of $\sim 400$, and resulted in values of 2200\,s, 1750\,s, and 300\,s for 
IRAS\,13258, IRAS\,06338, and HD\,52961, respectively.
The spectra were extracted from the droop and RSC data products, processed through the S18.0
version of the SSC (Spitzer Science Centre) data pipeline.
For the spectral extraction we used
data reduction packages, developed for the c2d (Core to discs) and feps (Formation and evolution of planetary systems) key projects. 
For a detailed description of these reduction packages, we refer to \citet{lahuis06} and \citet{hines05}.
The resulting infrared spectra for our sample stars can be seen in Figure~\ref{3stars}.
As shown in \citet{gielen11}, the infrared spectra of these three stars show emission features
which can not be explained by the typical silicate dust species, as is the case for 
the rest of the larger sample post-AGB disc sources. Details for the programme stars
can be found in Table~\ref{3stars}.

For IRAS\,06338+5333 (or HD\,46703, IRAS\,06338 hereafter) and HD\,52961 the binarity is confirmed, and orbital parameters are given in Table~\ref{sterren}.
These two stars also show the depletion pattern of condesable elements in the stellar photosphere \citep{waelkens91b,hrivnak08}, commonly seen in post-AGB disc sources \citep[e.g.][]{giridhar05,maas05}.
This abundance pattern, characterised by a lack of refractory elements, is the result of gas-dust separation in the disc,
followed by a re-accretion of the cleaned gas component \citep{waters92}. HD\,52961 is an extreme example of this process, having a [Fe/H] ratio of $-4.8$ and [Zn/Fe] = $+3.1$.  Chemical studies have shown that the stars are oxygen-rich, with C/O$<1$ \citep{luck84,bond87,waelkens91b,vanwinckel92,bakker97,hrivnak08}.  
Unfortunately, no C/O ratio has been determined for IRAS\,13258-8103 
(IRAS\,13258 hereafter). Optical spectra obtained so far cannot give conclusive evidence for either 
a carbon- or oxygen-rich nature of the stellar photosphere (Thomas Lloyd Evans, private communication).
So far, all post-AGB disc sources where we have studied the stellar photosphere, show an O-rich chemistry \citep[e.g.][]{gonzalez97a,giridhar98,giridhar00,vanwinckel03}.
So it is most likely that this star also follows this trend. If not, and the photosphere is C-rich in stead, this would
be the first C-rich post-AGB disc source detected.

All three sample stars show the typical broad infrared excess, associated with the presence of a circumbinary disc \citep{deruyter06,giridhar05}.
This excess stars already at near-infrared wavelengths, and is very different from the excess related to an outflowing shell,
where the excess starts and peaks at much longer wavelengths.
However, HD\,52961 and IRAS\,06338 do not show the very high $L_{IR}/L_*$ ratio, 
seen in other post-AGB disc sources \citep[see Fig.~1 in][]{gielen11}.
For HD\,52961 $L_{IR}/L_*=12\%$ and IRAS\,06338 has a ratio of only 3\%, whereas the typical value for 
such sources is around 40-50\% \citep{gielen08,gielen11}.
Although for IRAS\,13258 the lack of stellar parameters and value for the
total reddening makes an exact determination of the $L_{IR}/L_*$ ratio difficult,
the SED shows that the infrared excess is significant \citep[$> 50\%$,][]{gielen11}.
\citet{lloydevans97} deduced that  for this star the disc is most likely seen very edge on, 
since the strong Na D emission lines in the optical spectrum show that the star is seen by reflection on 
the circumstellar matter, which points to a close to edge-on orientation of the disc.

IRAS\,06338 and HD\,52961 have strong similarities with another post-AGB disc source,
namely EP\,Lyr. 
This stars also has an O-rich classification \citep{gonzalez97a}, and shows strong class C PAH features and CO$_2$ gas line emission (see Fig.~\ref{3stars}), 
together with strong features due to crystalline silicates \citep{gielen09}. Furthermore, EP\,Lyr also has an
atypically low $L_{IR}/L_*$ ratio of 3\%.

\section{Silicate dust features}
\label{sect_silicates}

In order to study the carbonaceous molecules visible in the spectra, we need to remove the underlying silicate dust contribution.
Our analysis of the residual carbonaceous emission is then further discussed in Sections~\ref{sect_pahs} and \ref{sect_c60}.

\begin{figure}
\vspace{0cm}
\hspace{0.cm}
\resizebox{9cm}{!}{ \includegraphics{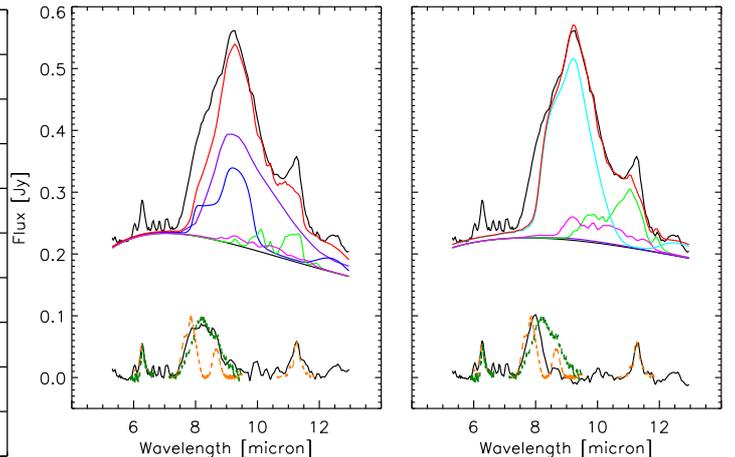}}
\caption{Results of our modelling of the 10\,$\mu$m feature in IRAS\,06338, using silica (model A, left) and alumino-silicates (model B, right). The different dust contributions are: amorphous pyroxene (violet), amorphous silica (blue), enstatite (magenta), forsterite (light green) and NaAlSi$_4$O$_{10}$ (cyan). The combined spectrum is given in red. On the bottom we show the residuals of this model, and overplot class B (orange) and class C (dark green) PAH emission features as described by \citet{peeters02} and \citet{vandiedenhoven04}.}
\label{silica_nasil}%
\end{figure}

\subsection{The 10\micron region}

The spectrum of IRAS\,06338 is dominated by a broad emission feature peaking at 9.2\,$\mu$m (see Fig.~\ref{silica_nasil}). 
The emission is clearly different from the typical 9.8\,$\mu$m feature seen in other post-AGB disc sources \citep[see Fig.~6 in][]{gielen11}, where the feature is mainly due to emission of amorphous and crystalline olivine.
Clearly, other dust species, which peak at shorter wavelengths, are needed to explain the observed emission.
One possibility is silica, which has different polymorphs that all peak around 9\,$\mu$m \citep{sargent09}. 
To determine the different contributing dust species to this feature, we assume the emission to come from an
optically thin region and linearly add dust absorption profiles, as done in \citet{gielen08,gielen11}. 
The mass absorption coefficients for the dust species are calculated 
from refractory indices in gaussian random fields (GRF) dust approximation \citep{shkuratov05}. 
The refractive indices for the different dust species are taken from \citet{servoin73,dorschner95,henning97}, and \citet{jaeger98}.
Detailed results of the modelling can be found in Table~\ref{fitresults}.

We can reproduce the feature using amorphous pyroxene, amorphous silica and forsterite (Fig.~\ref{silica_nasil}: Left). The residual emission at $6.04-6.28-8.23$ and 11.28\,$\mu$m can be explained by emission due to PAHs. Note that we do not find evidence for
the presence of amorphous olivine, even though it is usually one of the most abundant species seen in the spectra of
post-AGB disc sources \citep{gielen08,gielen11}, and we do see features due to crystalline olivine at longer wavelengths (see Sect.~\ref{longwaveregion}).
Some alumino-silicates also peak around 9\,$\mu$m \citep{mutschke98}, and a good fit can also be obtained using NaAlSi$_4$O$_{10}$, forsterite and enstatite (Fig.~\ref{silica_nasil}: Right). The residual spectrum is then very similar to the former case, but the 8.23\,$\mu$m residual feature is now blueshifted to 8.00\,$\mu$m, with an additional feature at 8.66\,$\mu$m. Again, the residual features can be explained by emission due to PAHs.
In none of the models, adding amorphous olivine improved the fit. The models described here are most likely not the only solutions,
since emission features of amorphous dust are often relatively interchangeable. However, all models show strong residual emission 
at 8\,$\mu$m, but the exact shape remains unclear.

\begin{figure}
\vspace{0cm}
\hspace{0.5cm}
\resizebox{8cm}{!}{ \includegraphics{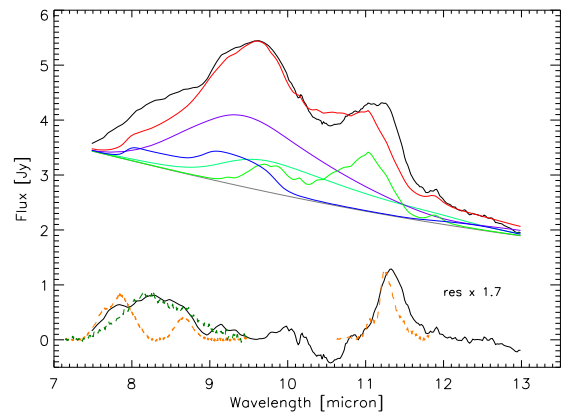}}
\caption{Results of our modelling of the 10\,$\mu$m feature in HD\,52961. The different dust contributions are: amorphous pyroxene (violet),
amorphous olivine (turquoise), amorphous silica (blue) and forsterite (light green). The combined spectrum is given in red,
with the continuum plotted in gray. On the bottom we show the scaled residuals of this model, and overplot  an example of class B (orange) and class C (dark green) PAH emission features.}
\label{hd52961_pahs}%
\end{figure}

HD\,52961 also shows a prominent feature in the 10\micron region, peaking around 9.7\,$\mu$m,
pointing to the presence of amorphous olivine and pyroxene. Again we see a clear 
shoulder at 8\micron and a strong feature at 11.3\,$\mu$m. This feature is much broader than what is observed
for IRAS\,06338 and is most likely a blend of forsterite and PAH emission.
We are able to get a reasonable fit to the spectrum using amorphous olivine/pyroxene, amorphous silica and forsterite (see Fig.~\ref{hd52961_pahs}). However, it proves difficult to reproduce the strong 11.3\micron without getting too much flux in the 
10.5\micron region. This could be due to the adopted synthetic spectrum for forsterite. The synthetic spectra of crystalline silicates
are known to be very sensitive to the exact chemical composition, temperature, the grain shape and grain size \citep[e.g.][]{koike03,koike10}.
Similarly, also the very strong observed 16\micron feature, generally associated with forsterite emission, proved difficult
to reproduce with current forsterite synthetic spectra \citep{gielen09,gielen11}. 
As for IRAS\,06338, we again find residual emission at wavelengths corresponding to PAH emission features.

IRAS\,13258 shows no strong silicate features in the 10\micron region. A weak feature can be seen 
ranging from 8 to 13\,$\mu$m. This could be due to weak emission of amorphous silicates.
Since there is evidence that we see the disc close to edge on, this could be due to heavy reddening of the central star and the hot
inner regions of the disc, where one would expect most of the 10\micron emission to arise from.

\subsection{The 18-36\micron region}
\label{longwaveregion}

\begin{figure}
\vspace{0cm}
\hspace{0cm}
\resizebox{9cm}{!}{ \includegraphics{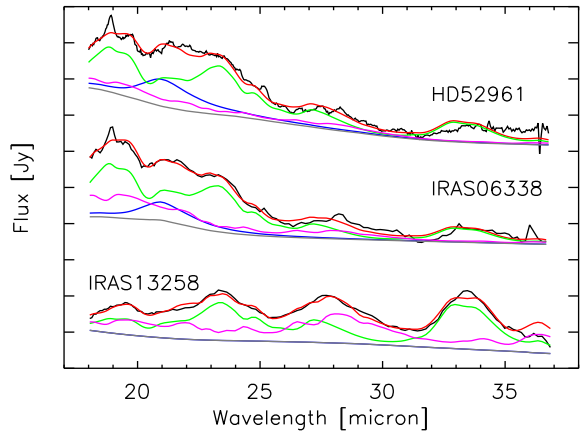}}
\caption{The long-wavelength Spitzer spectrum of our sample stars, together with the model spectrum (red)
consisting of amorphous silica (blue), forsterite (light green) and enstatite (magenta). The adopted continuum is given in gray.
The spectra are scaled and offset for comparison.}
\label{longwave}%
\end{figure}

For all three stars, the observed long-wavelength spectra show emission features that can be reproduced by emission of amorphous olivine/pyroxene,
forsterite/enstatite and amorphous silica. In Figure~\ref{longwave} we plot the observed long-wavelength spectra, together with model spectra consisting of the above mentioned dust species. 
Detailed results of the modelling can be found in Table~\ref{fitresults}.
Since the limited wavelength range makes it difficult to distinguish the underlying continuum from contribution of amorphous olivine/pyroxene, we only include amorphous silica, forsterite and enstatite in our modelling.
NaAlSi$_4$O$_{10}$, which we used in the 10\micron modelling of IRAS\,06338, has a similar feature as silica at 21\,$\mu$m, and so both dust species are interchangeable in this model. The features at 21\micron of NaAlSi$_4$O$_{10}$ and SiO$_2$ are relatively weak compared to 
the 10\micron feature, especially at higher temperatures, so we do not expect to see a very strong contribution of these species at 21\,$\mu$m.

For IRAS\,06338 and HD\,52961, forsterite is clearly the dominant crystalline species, with a smaller contribution of
enstatite and amorhous silica. In IRAS\,13258, there is a 
significant contribution of enstatite, but there is no evidence for amorphous silica.
For IRAS\,13258, the strong forsterite 33.6\,$\mu$m feature points to the presence of a large fraction of cool crystalline silicates in the disc.
If we see the disc very edge on, we might see more of the cooler 
outlying material, explaining the dominance of the cool long-wavelength features in the spectrum.

\section{CO$_2$}
\label{sect_co2}

\begin{figure}
\vspace{0cm}
\hspace{0cm}
\resizebox{9cm}{!}{ \includegraphics{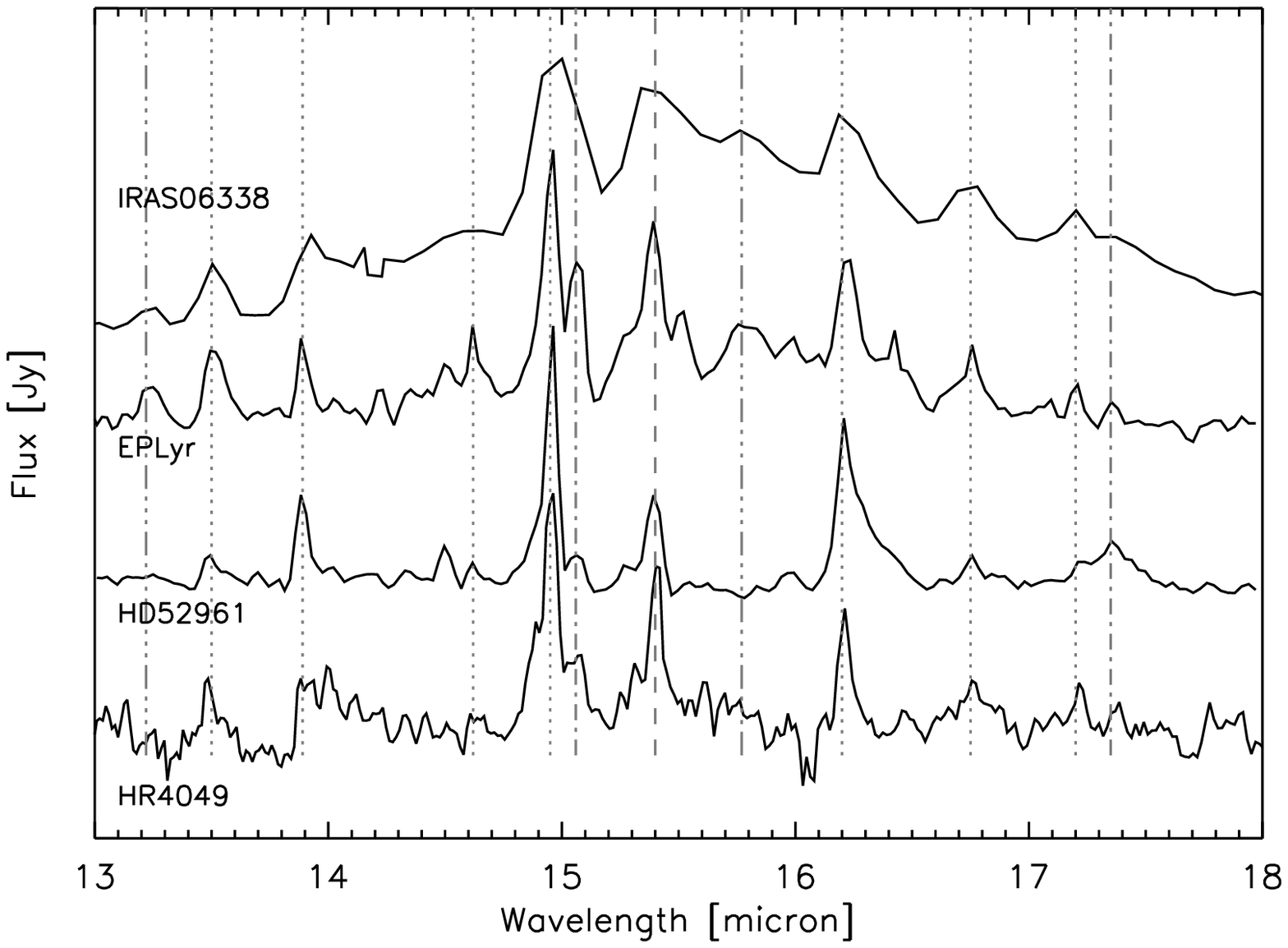}}
\caption{The CO$_2$ emission region of IRAS\,06338, compared to other
  post-AGB sources showing CO$_2$ gas emission
  \citep{cami01,gielen09}.  Dotted lines show the location of
  $^{12}$C$^{16}$O$_2$ features, dashed lines give
  $^{13}$C$^{16}$O$_2$ and dot-dashed lines show
  $^{16}$O$^{12}$C$^{18}$O.  Dot-dot-dashed lines mark unidentified
  features.  The spectra are scaled and offset for comparison. For HD\,52961, the underlying
  feature at 16\micron was removed.}
\label{co2}%
\end{figure}

In IRAS\,06338 and HD\,52961 we see strong narrow lines in the $13-17$\micron region,
which can be identified with CO$_2$ gas emission (Fig.~\ref{co2}).

For HD\,52961, an underlying feature at 16\,$\mu$m is seen,
probably (partly) coming from crystalline silicates, more specifically forsterite. 
The CO$_2$ gas emission in IRAS\,06338 is similar to the emission seen in HD\,52961 and EP\,Lyr \citet{gielen09},
and can also be identified with the CO$_2$ isotopes $^{12}$C$^{16}$O$_2$, $^{13}$C$^{16}$O$_2$ and
$^{16}$O$^{12}$C$^{18}$O.
Unfortunately, the low spectral resolution of IRAS\,06338 does not allow 
to determine the exact isotope abundances, as was done for HD\,52961.

The CO$_2$ lines in IRAS\,06388 seem to by lying on top a very broad ($\sim 4$\micron) feature at 15\,$\mu$m.
Note that the 15\micron feature observed in IRAS\,06338 is much
broader than the 16\micron feature we see in HD\,52961, and thus
probably does not have the same origin. Of course, this 16\micron
feature could contribute to the observed feature in IRAS\,06338, but
additional emission is still needed to give rise to such broad
emission.  Only a few dust species have strong emission features in
the 15\micron region, most of them are titanium-oxide combination, but none
of those species provides an explanation for the observed feature.

A likely explanation for the origin of this broad emission
band is that there is a large column of optically thick CO$_2$ gas. 
Simple isothermal LTE models of optically
thick CO$_2$ gas show a broad and strong plateau underneath the
strong Q-branch bands. This CO$_2$ plateau is made up by the
multitudes of ro-vibrational lines that bunch up in this wavelength
range. Such plateaus can indeed reproduce the overall width of the
observed emission plateau, but overestimate the flux shortward of 15\,$\mu$m. 
However, if the gas is optically thick, the spectrum can be
very sensitive to temperature and density gradients in the gas, and
it is thus conceivable that a more adequate model would in fact
reproduce the observations quite well. Such is also the case for HR\,4049, 
where optically thick models can simultaneously explain the
broad emission plateau and the individual strong Q-branch bands
(Malek et al., in preparation).

\section{PAHs}
\label{sect_pahs}

After the removal of the contribution of the O-rich dust (Sect.~\ref{sect_silicates}), we find clear evidence for PAH features in all sample stars.
The PAH emission features are generally divided in three classes, depending on the central wavelength of the most prominent features
\citep{peeters02}.
“Class A” sources have features at 6.22, 7.6 and 8.6\,$\mu$m. “Class B”
sources show quite some variability in peak positions, but tend to
be shifted more to the red than class A features. The more rare “class
C” sources show emission features at 6.3\,$\mu$m, no emission near
7.6\,$\mu$m, and a broad feature centred around 8.2\,$\mu$m, extending
beyond 9\,$\mu$m.

As already discussed in Section~\ref{sect_silicates}, the exact
identification of the PAH emission in the 8\,$\mu$m region for
IRAS\,06338 is uncertain. The features at $6.04-6.28-11.28$ and
12.6\,$\mu$m are relatively independent from the underlying dust
emission, unlike the 8\,$\mu$m feature, where the peak position can
shift from 8 to 8.3\,$\mu$m, depending on the composition of the
underlying silicate dust (see Fig.~\ref{silica_nasil}). When peaking
at 8\,$\mu$m, the feature resembles more a class B PAH
feature. With the peak at 8.3\,$\mu$m, the feature is identified
as class C.

IRAS\,06338 also shows very unusual emission in the 7\,$\mu$m
region. Three clear features at 6.63, 6.83 and 7.07\,$\mu$m can be
(see inset Fig.~\ref{3stars}), together with two smaller features at
6.49 and 7.25\,$\mu$m. 
The features are reminiscent of the narrow features
seen in the PAH spectra of \citet[][Fig.~2]{cami11}, where the
features are due small PAHs, consisting of less than 30 carbon atoms.
Similar features at 6.85 and 7.25\,$\mu$m have been
observed in a few other sources and identified with
aliphatic hydrocarbons and/or hydrogenated amorphous carbon
\citep{furton99,chiar00,sloan07}. A small feature at $6.6-6.7$\micron has been observed in
other circumstellar environments, and tentatively identified with PAH emission
\citep{peeters99,werner04b,smith07}.

\begin{figure}
\vspace{0cm}
\hspace{0cm}
\resizebox{8cm}{!}{ \includegraphics{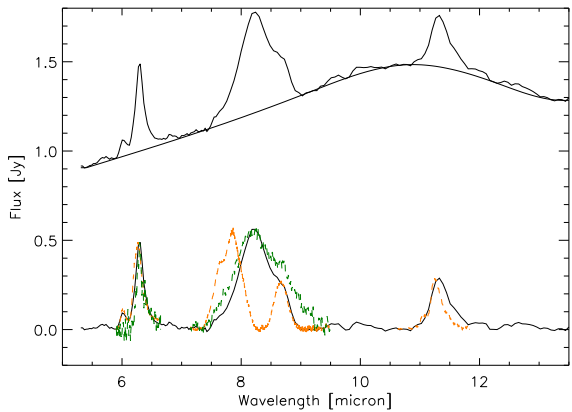}}
\caption{Identification of the PAH features in IRAS\,13258. On the top
  we plot the spectrum shortwards of 13\,$\mu$m, together with a
  spline fit to represent the underlying continuum. The residual
  spectrum is given below, together with an example of the class B
  (orange) and class C (dark green) PAH features as described by
  \citet{peeters02} and \citet{vandiedenhoven04}.}
\label{pahs_IRAS13258}
\end{figure}

The spectrum of IRAS\,13258, shortward of 15\,$\mu$m, is dominated by
PAH emission, with features at 6.02, 6.30, 8.22 and 11.30\,$\mu$m.
The large PAH feature around 8\,$\mu$m is again relatively broad, and
appears to be a blend of smaller features, with peaks at 7.65, 8.22
and 8.58\,$\mu$m. This broad band resembles the class C profile
  but is clearly narrower and has a red wing that resembles more a
  highly redshifted B profile (peaking at $~8\,\mu$m). Given that the peak
  position is 8.22\,$\mu$m, we classify it as a C profile. Other small
peaks can be seen at 9.1, 9.6, 10.0, 10.6 and 12.6\,$\mu$m. The
spectrum shows strong similarities with that of EP\,Lyr (see Fig.~\ref{3stars}), which also
shows PAHs emission at shorter wavelengths and strong crystalline
silicate emission at longer wavelengths \citep{gielen09}.

For HD\,52961 the residual spectrum (Fig.~\ref{hd52961_pahs}) shows
similar PAH features as for the previous two sources.  We again find a
broad 8\micron feature, with a possible small side feature at
9.2\micron. This would then make it a BC class \citep{sloan05}. The 11.3\micron feature is very similar to that observed
in IRAS\,13258, with a similar width and central wavelength.

\section{Fullerenes: C$_{60}$}
\label{sect_c60}

As can be seen already from Figure.~\ref{3stars}, the spectra of HD\,52961 and IRAS\,06338 
show weak features near 17.4 and 19\,$\mu$m (Fig~\ref{hd_iras_c60}). These features have
recently been identified with the infrared active vibrational modes of
neutral C$_{60}$ in the Spitzer spectrum of the C-rich planetary nebula Tc~1 \citep{cami10}. 
In that source, the bands show fairly broad emission bands with a roughly Gaussian profile,
similar as in our objects. However, in our spectra, the short-wavelength bands at 
7 and 8.5\,$\mu$m are missing, as are bands of C$_{70}$.

The two C$_{60}$ features are weak, and especially the 17.4\,$\mu$m
band is furthermore contaminated by some CO$_2$ emission (see Sect.~\ref{sect_co2}). 
We can get an estimate of the physical conditions and masses involved, following
a similar approach as described in \citet{cami10}, using the emitted power in each of the bands.  Given the
absence of the bands at shorter wavelengths and contamination with CO$_2$ lines (see Sect.~\ref{co2}), the diagnostic value of
the C$_{60}$ bands is however somewhat limited. 

\subsection{HD\,52961}

\begin{figure}[h]
\vspace{0cm}
\hspace{0cm}
\resizebox{9cm}{!}{ \includegraphics{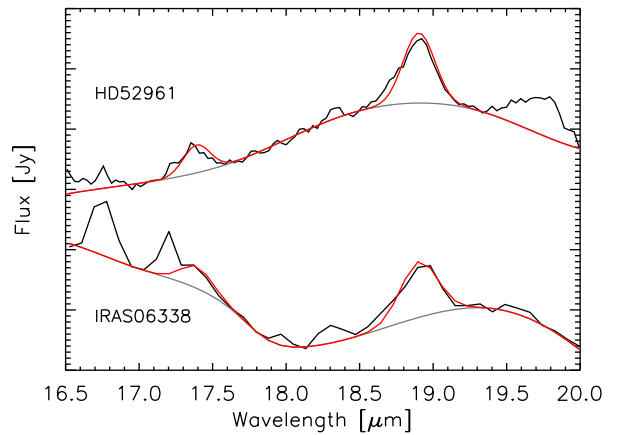}}
\caption{The observed spectra of HD\,52961 and IRAS\,06338 (black) with the estimated spline continuum (grey).
In red we overplot the Gaussian profiles that were fit to the two C$_{60}$ bands. The spectra are scaled and offset for comparison.}
\label{hd_iras_c60}%
\end{figure}

The band profiles of the C$_{60}$ bands in HD\,52961 appear fairly
symmetric, and Figure~\ref{hd_iras_c60} shows that a gaussian profile reproduces the
observations well. From integrating
the spectrum, we find that the total power emitted in the C$_{60}$
bands is $2.6 \pm 0.2 \times 10^{-16}$ and $6.5 \pm 0.2 \times
10^{-16}$\,Wm$^{-2}$ for the 17.4 and 18.9\,$\mu$m bands
respectively. From the ratio of these numbers, we then find a nominal
excitation temperature of 152$^{+29}_{-22}$\,K . At a distance of 2.1\,kpc, the observed emission
requires a total mass of about $3.3 \times 10^{-8}$\,M$_{\odot}$. Note that these uncertainties do not include
estimates of systematic errors made by propositioning of the continuum 
level.

The observed widths of the features in HD\,52961 are somewhat narrower
than in the C-rich planetary nebula Tc~1 \citep{cami10}. Smaller widths are expected for lower
temperatures, and indeed, also the excitation temperature for HD\,52961
is somewhat lower than the $\sim$330\,K found for Tc~1. Note though
that the observed widths are smaller than what would be expected based
on laboratory experiments where band widths have been measured at
various temperatures \citep[e.g.][]{nemes94}. The low temperature also
explains the absence of the two bands at shorter wavelengths: at an
excitation temperature of 200\,K and assuming thermal excitation, the
total power in each of the bands at shorter wavelengths is less than
10\% of the power in the 17.4\,$\mu$m band.

\subsection{IRAS\,06338}

The IRAS\,06338 observations correspond to a much lower spectral
resolution than those of HD\,52961, and at the same time the C$_{60}$
features are weaker, and heavily contaminated by high-excitation
emission bands due to CO$_2$. This makes a good determination of the
relevant parameters much harder, and any derived quantity is probably
rather uncertain given the systematic errors possible on the local
continuum. We masked the CO$_2$ bands, and adopted a local continuum,
and fitted Gaussian profiles to the bands (See Fig.~\ref{hd_iras_c60}).

The widths (FWHM) of the 17.4 and 18.9\,$\mu$m bands are comparable to what we find for
HD\,52961. The emitted power is $2.7 \times \ten{-17}$ and $7.0 \times \ten{-17}$\,Wm$^{-2}$
respectively, yielding a nominal excitation temperature of 145\,K and
a C$_{60}$ mass of $1.5 \times \ten{-8}$\,\Msun (adopting a distance of 3.9\,kpc).

\section{Discussion}
\label{sect_discussion}

\subsection{Mineralogy}

In all three sample sources, we find evidence for O-rich silicate dust features.
Both IRAS\,06338 and HD\,52961 have a peculiar silicate mineralogy, very different to what is generally 
seen in post-AGB disc sources \citep{gielen11}.
The sources seem to be dominated by warmer dust species, with strong emission in the 10\micron region, but very weak features
at longer wavelengths. This is also seen in interferometric measurements of HD\,52961, where the crystalline grains seem to be concentrated in the
inner disc regions \citep{deroo06}.
The peak position of the 10\micron feature in IRAS\,06338 points to a lack of amorphous olivine, with a preference for amorphous pyroxene and silica, or alumino-silicates, in stead. However, both at short and long wavelengths we find clear features of forsterite, the crystalline olivine form.
HD\,52961 shows a 10\micron feature which is considerably less broad than other post-AGB disc sources \citep{gielen11}, 
again pointing to the dominance of amorphous pyroxene and silica over amorphous olivine.
The star also has unusually strong features at 11.3 and 16\,$\mu$m, wavelengths typically associated with forsterite emission. 
The exact shape and strength of these features proved difficult to model with current synthetic spectra of forsterite.

The underabundance of amorphous olivine and weak crystalline silicate features in IRAS\,06338 and HD\,52961
show that the dust is rather metal-poor. 
It remains unclear whether this peculiar mineralogy could be linked to the extreme chemistry in the stellar photosphere. 
The photosphere for HD\,52961 shows a strong deficiency of metals, with [Fe/H] $=-4.8$ \citep{waelkens91b}, which are currently locked in dust grains in the disc. 
IRAS\,06338, however, does not show such extreme depletion, rather its value of [Fe/H] $=-1.6$ and [Mg/H] $=-1.5$ \citep{hrivnak08} are average values found for post-AGB disc sources \citep{deruyter06}. Furthermore, HR\,4049 has a similar extreme depletion pattern as HD\,52961 \citep{waelkens91a}, but has a very different infrared spectral signature, with features only due to CO$_2$ gas and class B PAHs, and no silicate emission. Unfortunately, due to this strong depletion pattern, it is very difficult to determine the initial metallicity of the 
central star.

\subsection{PAHs}

The PAH emission seen in all sample stars can be identified as class C
(or BC), depending on the underlying silicate emission \citep[][and references therein]{peeters11}. All our sources follow the trend
between the effective temperature of the central star and the central
wavelength of the $7-8$\micron PAH feature
\citep{sloan07,smolders10}. 
The carriers of the class C features are proposed to be aliphatic carbonaceous
material \citep{sloan07}. If indeed aliphatic carbonaceous material is responsible for the
class C profile, we would expect more aliphatic bands at 6.85 and 7.25\,$\mu$m,
which are (tentatively) observed in IRAS\,06338 and IRAS\,13258.

\subsection{Fullerenes}

Even though the diagnostic value of the C$_{60}$ bands in 
IRAS\,06338 and HD\,52961 is rather limited,
the derived low excitation temperatures show that the fullerenes cannot
be located very close to the central star. 
Adopting a luminosity of 5000\,L$_{\odot}$, a dust temperature of 150\,K 
corresponds to a distance of about 240\,AU.
Unfortunately, neither for the PAH or fullerene features, we can
determine their exact spatial location.

Several possible mechanisms for the formation of fullerenes in the CE
of evolved stars are discussed by \citet{cami11b}. The classic
experiments that led to the discovery of fullerenes \citep{kroto85} 
show that fullerenes self-assemble from collisions between
carbon clusters in a H-poor environment. The presence of H under
similar conditions inhibits the formation of fullerenes, and instead
results in the formation of PAHs \citep{kroto88,devries93}.
At much higher temperatures ($>3000$\,K), 
fullerenes do form in the presence of hydrogen \citep{jaeger09}. 
None of these mechanisms seem to correspond to the
conditions that could be expected in the outflows of evolved stars
though. The simultaneous presence of fullerenes and PAHs in several
PNe led \citet{garciahernandez10} to conclude that PAHs and
fullerenes form from photoprocessing of hydrogenated amorphous carbon
(HAC), as was experimentally suggested by e.g. \citet{scott97}.
However, if this would be the fullerene formation process
in PNe, it is hard to understand why fullerenes are only detected
in low-excitation PNe, and it would be even more unlikely to
invoke such a process for the formation of fullerenes in our post-AGB
sources, where the UV irradiation is considerably weaker. One of the
more promising avenues is the more recent study of \citet{micelotta10}, 
that shows fullerenes could form from the destruction of PAHs due to shocks.
\citet{micelotta10} find that shocks with
velocities between 75 and 100\,km/s are needed to modify PAH molecules in the
ISM. For several post-AGB disc sources, shocks due to strong winds
(reaching speeds well above 100\,km/s) and pulsation (due to the
crossing of the Cepheid instability strip) have been measured
\citep[e.g.][]{gillet89,vanwinckel98,maas03,vanwinckel03,maas05}. Thus, this scenario could well be
applicable for the shocked material around these post-AGB binaries, where
the higher material densities involved could be beneficiary to the fullerene formation.

\begin{table*}
\caption{Overview of the different characteristics of post-AGB disc sources with evidence
for carbonaceous molecules from the list given by \citet{deruyter06} and \citet{gielen11}. We list the name, spectral type, effective temperature, C/O ratio, the $L_{IR}/L_*$ ratio,
the presence of amorphous and/or crystalline silicates, the detection of PAHs and/or C$_{60}$,
and the presence of CO$_2$ gas lines in their infrared spectra. }
\label{starcharac}
\centering
\begin{tabular}{lcccrccccc}
\hline \hline   
 	Name			 & Spec. Type & $T_{\rm eff}$ (K) & C/O  & $L_{IR}/L_*$ & Am. Sil. & Cryst. Sil. & PAHs & C$_{60}$ & CO$_2$\\
\hline
\\
EP\,Lyr & A4I & 7000 & $< 1$ & 3\% & no/weak & strong & class C & no & yes\\
IRAS\,06338 & F3I & 6250 & $<1$ & 3\% & yes & weak & class (B)C & yes & yes \\
HD\,52961 & F6I & 6000 & $<1$ & 12\% & yes & weak & class C & yes & yes \\
HR\,4049 & A6I & 7500 & $< 1$ &  25\% & no & no & class B & no & yes\\
IRAS\,13258 & F4Ib-G0Ib & ? & ? & $>50\%$ & ? & strong & class C & no & no \\
Red Rectangle & F1I & 7500 & $\leq 1$ & $> 100\%$ & no & strong & class B & no & no\\ 
\hline
\end{tabular}
\end{table*}

\subsection{Carbonaceous molecules in an O-rich environment}

Given the O-rich classification of IRAS\,06338, HD\,52961 and EP\,Lyr, there must be a process that 
can free some of the locked carbon in the CE to form the observed large
carbonaceous molecules. 
CO molecules can be destroyed due to photo-dissociation, but the UV radiation 
of these sources might not be strong enough to dissociate large quantities of CO. 
Alternatively, X-ray radiation, caused by shocks, is able to break up CO \citep{woods05}.
Another possibility is the Fischer-Tropsch catalysis mechanism,
a chemical reaction process that can form hydrocarbons and H$_2$O out of CO and H$_2$.
This mechanism has been invoked to explain the presence of C-rich molecules in other
O-rich environments and the solar nebula \citep{kress01,jura06}. 
PAHs have also been found around O-rich PNe central stars \citep[e.g.][]{guzmanramirez11},
usually surrounded by a dense torus.
They present a chemical model that shows that hydrocarbon chains can form within
an O-rich environment, but high UV radiation is necessary to break up CO,
which is not present in our post-AGB disc sources.

Another scenario could be that the central stars have become C-rich, 
but that due to the reaccretion of O-rich gas from the circumbinary disc,
an optically thick layer is newly formed on the stellar photosphere \citep{mathis92}.
The effects of this accretion process can be seen in the photospheric
depletion of metals, but it is unclear whether the reaccreted gas is 
rich enough in oxygen, or the accretion rate high enough, to explain the apparent O-rich stellar photosphere.
Furthermore, we know that post-AGB stars often undergo strong pulsations \citep{vanwinckel99,vanwinckel09},
which might easily mix the different layers of the stellar photosphere.
For IRAS\,06338 and HD\,52961 pulsation periods of, respectively, 29 and 72 days are found \citep{waelkens91b,hrivnak08}.

In the case of IRAS\,13258 where the C/O ratio is not known, and the
central sources could thus be carbon rich, the carbonaceous molecules
can be located in a more recent, C-rich outflow, while the silicate dust
is located in the O-rich disc.

\subsection{Relation to other characteristics}

The question remains why C-rich molecules are observed in only a few
of the larger sample of $\sim 70$ post-AGB disc sources as given in \citet{deruyter06} and \citet{gielen11}. 
An overview of the different observational characteristics of these six sources is given in Table~\ref{starcharac}.

So far, most of the post-AGB disc sources where PAH emission is seen, also show mid-infrared
CO$_2$ gas emission; IRAS\,13256 and the Red Rectangle being the exception. No CO$_2$ gas lines are seen in other sources from
the larger Spitzer sample presented in \citet{gielen11}. This could mean that the CO$_2$ gas is linked to the formation
of carbonaceous molecules in an O-rich CE; or that an underlying mechanism gives rise to the different observed characteristics.
The presence of CO$_2$ gas could be a by-product of the Fischer-Tropsch (FT) catalysis.

Besides the CO$_2$ and PAH emission, IRAS\,06338, HD\,52961 and EP\,Lyr also all show unusually low $L_{IR}/L_*$ ratios (see Sect.~\ref{sect_progstars}).
So far, it is unclear whether the low $L_{IR}/L_*$ ratio in these sources is a result of a difference in disc formation
or a signature of disc evolution. The low infrared excess could be due to a lower dust mass in these discs,
compared to the larger sample. A lower dust-to-gas ratio might then make the gas lines more prominent,
explaining the observed strong CO$_2$ emission in these sources.
The low infrared excess might point to a disc structure, which is more 
transparent to stellar photons. This would then increase the
photodissociation rate of CO. 

\section{Conclusions}
\label{sect_conclusions}

So far, more than 60 infrared spectra of post-AGB disc sources, as listed by \citet{deruyter06}, have
been studied, both in our own Galaxy and the Large Magellanic Cloud \citep{waters98,dominik03,gielen08,gielen11}.
The spectra are nearly always dominated by emission due to
O-rich silicate features, with a high degree of crystalline dust.
Of this large sample, the total number of sources where we find evidence for
species besides silicate dust now comes to six, including the new objects discussed in this paper.

In these six sources the additional features are due to
O-rich CO$_2$ gas lines and/or carbonaceous molecules, most often PAHs.
Our detection of C$_{60}$ in HD\,52961 and IRAS\,06338 represents the
first time these molecules are found in binary post-AGB stars. The presence
of large aromatic molecules in these environments is very surprising
and puzzling, given the O-rich nature of the circumbinary disc,
and possibly the central star itself.

\bibliographystyle{aa}
\bibliography{referenties.bib}

\onecolumn
\begin{appendix} 

\section{Tables}
\begin{table*}[h]
\caption{Best fit parameters deduced from our full spectral fitting. The abundances of small (0.1\,$\mu$m) and large (2.0\,$\mu$m) grains (S-L) of the various dust species are given as fractions of the total mass, excluding the dust responsible for the continuum emission.}
\label{fitresults}
\centering
\begin{tabular}{lcccccccc}
\hline \hline   
 				 & MgFeOliv & MgFePyr  & MgOliv & MgPyr & Silica & Forst & Enst & NaAlSi$_4$O$_{10}$\\
             & S - L & S -  L & S - L & S -  L  & S - L & S -  L & S -  L & \\
\hline
\\
 IRAS\,06338 & &  & & & &&&\\
  10\micron region A    &$ 0.0   -   0.0$    & $68.3    -     0.0$&$ 0.0   -   0.0$ &$ 0.0   -   0.0$  & $ 0.0 - 21.3$ & $6.6 - 0.0$ & $1.4 - 2.4$  & -  \\
  10\micron region B & $0.0 - 0.0$ & $ 1.2 - 0.0$&$ 0.0   -   0.0$ &$ 0.0   -   0.0$  &  $0.0 -0.0$ & $11.6 - 21.1$ & $3.1 - 10.9$ & $51.4$ \\  
  20\micron region & - & - & - & - & $0.0 - 16.5$ & $15.7 - 41.3$ & $0.0 - 26.5$ & - \\\\
 HD\,52961 & &  & & & &&&\\
   10\micron region & $0.0 - 0.0$ & $ 0.0 - 0.0$ & $32.0 - 0.0$ & $ 37.6 - 0.0$&$ 0.0   -   10.0$ &$ 20.4   -   0.0$  &  $0.0 -0.0$ & - \\  
  20\micron region & - & - & - & - &$0.0 - 26.5$ & $17.8 - 40.2$ & $0.0 - 15.5$ & -\\\\  
 IRAS\,13258 & &  & & & &&&\\
  20\micron region & - & - & - & - & $0.0 - 0.0$ & $22.8 - 29.6$ & $9.5 - 38.2$ & -\\

\hline
\end{tabular}
\end{table*}

\end{appendix}

\end{document}